\documentstyle[epsf]{ptptex}

\title
{Molecular Structure with Exotic Clusters in light
Neutron-rich Nuclei}

\author{Y. Kanada-En'yo}

\inst{Institute of Particle and Nuclear Studies, \\
High Energy Accelerator Research Organization,\\
1-1 Oho, Tsukuba, Ibaraki 305-0801, Japan}

\abst{
The excited states of $^{12}$Be, $^{14}$Be and $^{15}$B have been 
studied with a method of antisymmetrized molecular dynamics.
In the predicted excited states we find novel molecule-like structures 
with very exotic clusters such as $^6$He+$^8$He in $^{14}$Be and $^6$He+$^9$Li 
in $^{15}$B.
The origin of the $^6$He cluster development in the very neutron-rich nuclei 
is understood by the new-type correlation among 4 
neutrons and 2 protons. In this paper we also present our recent 
challenge to study  $sd$-shell nuclei. Shape coexistence problems 
in $^{36}$Ar and $^{40}$Ca are discussed.}

\begin{document}

\maketitle

\section{Introduction}
In the light nuclear region, a clustering aspect 
is one of the essential features in unstable nuclei
as well as in stable nuclei.
Owing to the progress of the experimental techniques, 
information for the excited states of light unstable nuclei 
have been increased rapidly.
The exotic clustering of 
light unstable nuclei is one of the attractive subjects in the 
experimental and theoretical researches.

For example, theoretical studies on $^{10}$Be 
\cite{ENYOf,OGAWA,DOTE,ITAGAKI,ENYOg}
suggested molecule-like structures in the excited bands $K=1^-_1$ and
$K=0^+_2$.
The other candidates for the molecule-like states are 
the excited states of $^{12}$Be which have been recently discovered
in the break-up reactions
into $^6$He+$^6$He and $^8$He+$^4$He \cite{FREER}.
On the other hand, the structure of the ground state of $^{12}$Be has been 
interesting because there are some experimental data which indicate 
the vanishing of magic number 8 in $^{12}$Be \cite{TSUZUKIa,IWASAKI}.
However $^{12}$Be has not been studied enough by full microscopic 
calculations assuming no cluster core. It is also important to 
search for molecule-like states in other neutron-rich nuclei
in order to understand universal features of clustering in unstable nuclei. 
Our first aim is to study the states covering the 
ground and the excited ones of light neutron-rich nuclei with 
antisymmetrized molecular dynamics (AMD). We study the structures of $^{12}$Be
and challenge to discover possible molecule-like states with exotic 
clusters in heavier neutron-rich nuclei, $^{14}$Be and $^{15}$B.

We apply the microscopic method of antisymmetrized molecular dynamics(AMD)
which has already proved to be a very useful approach for the structure
study of general light nuclei \cite{ENYObc,ENYOe,ENYOg,ENYOsup}.
In the study of $^{12}$C \cite{ENYOe},
the author has proposed a new version of the method, variation after 
spin-parity projection(VAP) in the AMD framework which is very useful to
investigate excited states with various kinds of structures such as 
 spherical shell-model-like structures and clustering structures.
The author and her collaborators have 
succeeded to describe the structures of excited states of $^{10}$Be
\cite{ENYOf} with the VAP calculations in the framework of AMD.
AMD is the method that is very suitable to search for the possible
 exotic clusters in foreign nuclei because we do not need any 
model assumptions such as the existence of clusters.

As mentioned above, clustering is an important feature in very light nuclei.
However it is an open problem whether or not clustering effects 
are seen also in unstable $sd$-shell nuclei. 
I challenge to study structures of $sd$ nuclei with AMD method. 
In this paper, the recent hot 
subjects of shape coexistence problems in $^{36}$Ar and $^{40}$Ca
have been studied with AMD. Although they are stable ones,
the studies of stable nuclei are very essential to investigate 
the unstable $sd$-shell nuclei.

In this paper, the structures of the excited states of light
neutron-rich nuclei, $^{12}$Be, $^{14}$Be and
$^{15}$B are studied in Sec.\ref{sec:light}. 
We discuss the mechanism of the clustering 
development from the view point of correlation among 
nucleons in the single particle orbits.
In the study of heavier nuclei with AMD,
the shape coexistence problems in $^{36}$Ar and $^{40}$Ca are
investigated in Sec.\ref{sec:sd}.

\section{Formulation}\label{sec:formulation}

An AMD wave function 
is a Slater determinant of Gaussian wave packets;

\begin{eqnarray}
&\Phi_{AMD}={1 \over \sqrt{A!}}
{\cal A}\{\varphi_1,\varphi_2,\cdots,\varphi_A\},\\
&\varphi_i=\phi_{{{\bf Z}}_i}\chi_i\tau_i :\left\lbrace
\begin{array}{l}
\phi_{{{\bf Z}}_i}({\bf r}_j) \propto
\exp\left 
[-\nu\biggl({\bf r}_j-{{\bf Z}_i \over \sqrt \nu}\biggr)^2\right],
\label{eqn:single}\\
\chi_{i}=
\left(\begin{array}{l}
{1\over 2}+\xi_{i}\\
{1\over 2}-\xi_{i}
\end{array}\right).
\end{array}\right.
\end{eqnarray}
where the centers of Gaussians ${\bf Z}_i$'s are complex variational
parameters. $\chi_i$ is an intrinsic spin function represented by
$\xi_{i}$ which is also a variational parameter specifying the direction 
of the $i$-th intrinsic spin.
 $\tau_i$ is an isospin
function which is fixed to be up(proton) or down(neutron)
 in the present calculations.

In order to obtain wave function of an excited states of light nucleus, 
we vary the parameters ${\bf Z}_i$ and $\xi_{i}$($i=1\sim A$) to
minimize the energy expectation value for the 
parity and total angular momentum eigenstate (VAP calculations), 
\begin{equation}
{\langle P^J_{MK'}\Phi^\pm_{AMD}|H|P^J_{MK'}\Phi^\pm_{AMD}\rangle \over
\langle P^J_{MK'}\Phi^\pm_{AMD} |P^J_{MK'}\Phi^\pm_{AMD}\rangle },
\end{equation}
where the operator of total angular momentum projection $P^J_{MK'}$ is 
$\int d\Omega D^{J*}_{MK'}(\Omega)R(\Omega)$.
The integration for Euler angle $\Omega$ is calculated numerically.
We adopt the frictional cooling method 
\cite{ENYObc} to obtain the minimum energy states. 
Thus we can obtain the lowest state for a given spin parity $J^\pm$ with 
VAP calculations.
For higher excited states we perform the energy variation for the
orthogonal component to the lower states by superposing wave functions. 
More details of the AMD method for the excited states 
with the variation after spin-parity projection are described in Refs 
\cite{ENYOe,ENYOg}.
By making VAP calculations for the lowest and 
the higher excited states with various sets of total spin and parity
$\{J^\pm\}$,
we obtain a lot of AMD wave functions $\{\Phi_1,\cdots,\Phi_m\}$, 
which approximately describe the intrinsic states of the 
corresponding $J^\pm_n$ states. The number $m$ 
is the number of the considered states.
Final results are attained by diagonalizing 
a Hamiltonian matrix
$\langle P^{J\pm}_{MK'}\Phi_i|H|
P^{J\pm}_{MK''}\Phi_j\rangle$ ($i,j=1\sim m$).

For the structure study of heavier nuclei such as $^{36}$Ar and $^{40}$Ca,
we make variation after parity projection but no spin projection (VBP) 
with a constraint AMD instead of full VAP calculations
to save computational time. The adopted constraint in the present calculations
is that the expectation values of total oscillator 
quanta must equal to a given number as 
$\langle a a^\dagger \rangle= W$. After VBP calculations with a constraint,
we make spin projection to obtain a energy curve as a function of 
total oscillator quanta. The obtained states are superposed 
by diagonalizing Hamiltonian Matrix 
$\langle P^{J\pm}_{MK'}\Phi_i|H|P^{J\pm}_{MK''}\Phi_j\rangle$.

\section{Results of $^{12}$Be, $^{14}$Be and $^{15}$B}
\label{sec:light}

We apply the AMD method for the excited states of $^{12}$Be, $^{14}$Be
and $^{15}$B. The adopted interactions in this work are the central force of 
the modified Volkov No.1 with case 3 \cite{TOHSAKI},
the spin-orbit force of G3RS \cite{LS} and the Coulomb force.
The Majorana parameter used here is $m=0.65$, 
and the strength of G3RS force is chosen to be  
$u_1=-u_2=3700$ MeV. We choose a optimum width parameter 
$\nu$ for the Gaussians of the single particle wave functions of each nucleus. 
With these parameters, the parity inversion of the 
$^{11}$Be ground state can be reproduced. 
The binding energies of $^{12}$Be, $^{14}$Be and $^{15}$B are 
61.9 MeV, 59.7 MeV and 73.1 MeV, which are smaller than the experimental
data 68.65 MeV, 69.77 MeV, 88.19 MeV, respectively.
We have checked that the underestimation with the present interaction 
parameters can be improved easily by using smaller Majorana 
parameter as $m=0.61$. With this parameter set the binding energy
of $^{14}$Be is 67.7 MeV which well agrees to the experimental data.
We also find that the excitation energy of $0^+_2$ in $^{14}$Be
calculated with $m=0,61$ is 4.2 MeV which is almost as same as the one 
with $m=0.65$.

\begin{figure}%----------------------------------------
\begin{minipage}[b]{7cm}
\noindent
\epsfxsize=0.9\textwidth
\centerline{\epsffile{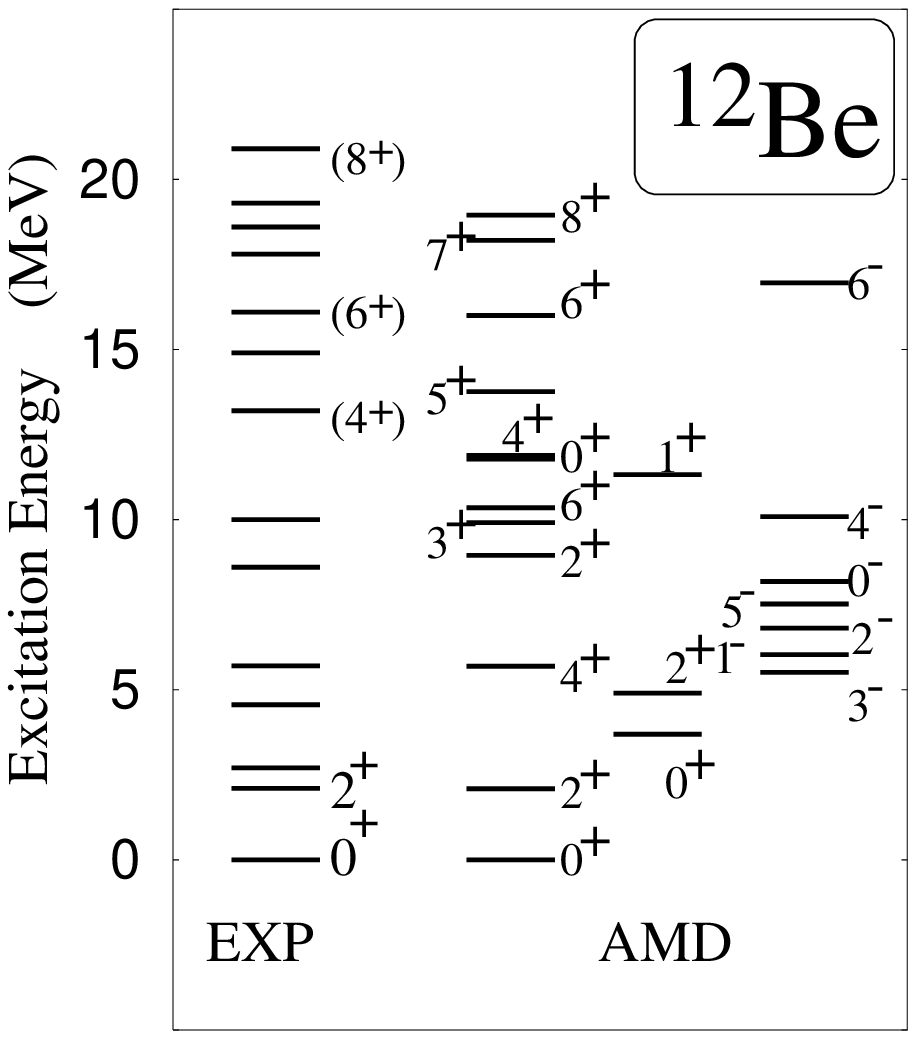}}
\caption{\label{fig:be12spe}
Excitation energies of the levels in $^{12}$Be. 
Theoretical results are calculated by VAP calculations based on AMD.
Experimental data are taken from the Table of Isotopes and Refs 
\protect{\cite{TANIHATA,FREER}}.
}
\end{minipage}\quad
\begin{minipage}[b]{7cm}
%\begin{figure}%----------------------------------------
\noindent
\epsfxsize=0.9\textwidth
\centerline{\epsffile{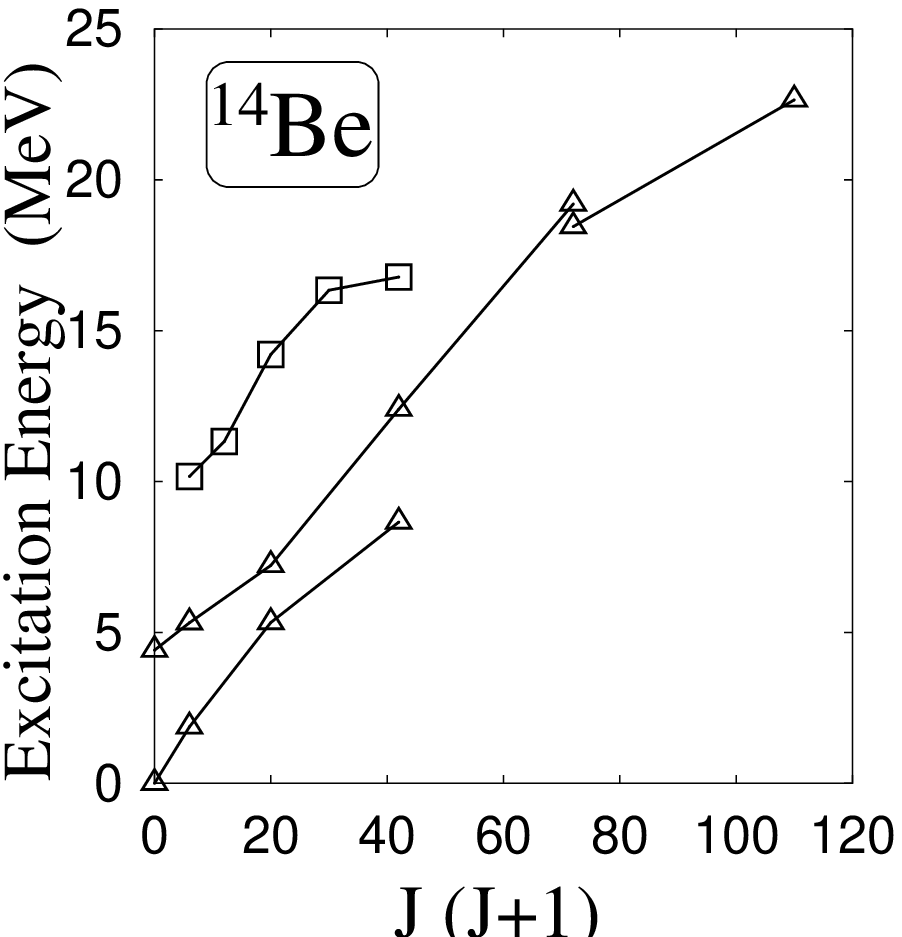}}
\caption{\label{fig:be14spe}
Excitation energies of the levels in $^{14}$Be
calculated by VAP calculations based on AMD.
Excitation energies are plotted as a function of $J(J+1)$
where $J$ is a total spin of the state.
}
%\end{figure}
\end{minipage}
\end{figure}

In the calculated results of $^{12}$Be with AMD, 
a lot of excited states appear in the low-energy region.
The energy levels of $^{12}$Be are presented in
Fig. \ref{fig:be12spe}. 
The theoretical levels of $4^+_2$, $6^+_2$ and $8^+_1$ states
well correspond to the recently 
observed excited states \cite{FREER}.
By analyzing the intrinsic AMD wave functions, we can classify 
the excited states into rotational bands such as 
$K^\pi=0^+_1,0^+_2,0^+_3,1^-_1$. The interesting point is that 
the newly observed levels \cite{FREER} at the energy region above 10 MeV 
have been found to 
belong to the third rotational band $K^\pm=0^+_3$.
It is also interesting to see the ground $K^\pm=0^+_1$ band 
properties because
the vanishing of the neutron magic number $N=8$ occurs in $^{12}$Be. 
Even though the $^{12}$Be is a neutron magic nucleus with $N=8$,
the intrinsic state of the ground $0^+$ state 
is not the ordinary state with the closed neutron $p$-shell, 
but a prolately deformed state with a developed 
clustering structure, which is dominated by $2p-2h$ configurations
in terms of single-particle orbits.
As a result, the ground $K^\pi=0^+$ band starts
from the ground $0^+$ state and reaches the band terminal at
$J^\pi=8^+$ with the highest spin state in the $2\hbar\omega$ configurations.
On the other hand, the neutron $p$-shell closed states 
with $0\hbar\omega$ configurations construct the second $K^\pi=0^+_2$
band which consists of the second $0^+$ and the second $2^+$
states. For an experimental evidence, the strength of the $\beta$ decay 
from $^{12}$Be(0$^+_1$) to $^{12}$B($1^+$) is helpful to estimate the
breaking of the neutron $p$-shell \cite{TSUZUKIa}.
As is expected, the theoretical value of B(GT)=0.8 
is enough small as the experimental data B(GT)=0.59 because 
the component of the $p$-shell broken state in the 
parent $^{12}$Be(0$^+_1$) makes the expectation value of Gamow-Teller
operator to be small. It is consistent 
with the pioneer works \cite{ITAGAKIa,TSUZUKIa}.
It should be pointed out that this is the first theoretical work 
which can reproduce systematically the energy levels from the 
ground state to the highly excited states.
We have predicted the new $K^\pi=0^+_3$ rotational band 
from $0^+_3$, $2^+_3$, $4^+_2$, $6^+_2$ and $8^+_1$ states. 
The $0^+_3$ state has the mostly developed $^{6}$He+$^6$He clustering. 
The developed clustering becomes weak gradually 
with the increase of the total-angular
momenta $J$, and finally it changes to the same spin aligned state 
at $J^\pm=8^+_1$ as the one in the $K=0^+_1$ band. 
The reason for the spin alignment at $J^\pm=8^+$ is because the
states in the  $K^\pm=0^+_3$ band are the other 
$2\hbar\omega$ states which must have the same highest spin state 
as the one in the ground band at the band terminal $8^+$.
Although the other evidence for the magic number vanishing is the
low-lying $1^-$ state which has been measured recently \cite{IWASAKI},
the excitation energy of $1^-$ state is overestimated in present calculation.
The mechanism of energy gain of negative parity states has been still 
an open problem.

By analyzing the single particle wave functions in the 
intrinsic state, the mechanism of the clustering development 
in the first $0^+$ state can be understood by an idea of molecular orbits
surrounding 2$\alpha$ cores. Although the molecular orbits in 
$^{12}$Be have been suggested theoretically 
\cite{OERTZEN,ITAGAKI},
the important point of the present work is that 
the structure of $2\alpha$ cores and surrounding neutrons in $^{12}$Be is
formed automatically in the energy variation though the existence of 
any clusters nor molecular orbits is not assumed.
On the other hand, it is natural to consider the third $0^+$ state as
$^6$He+$^6$He clustering structure instead of the $\alpha$ and surrounding
neutrons because the distance between two clusters is too far to be 
described in terms of the molecular orbits.
In this case, an $\alpha$ cluster goes outward
far enough to form a $^6$He cluster in the correlation with 2 valence
neutrons. 

Next we present the results of $^{14}$Be. 
Figure\ref{fig:be14spe} shows the energy levels of the positive parity 
states of $^{14}$Be.
While few excited states are known experimentally, 
many excited states are predicted in the theoretical results.
There are very recent experimental data \cite{SAITO}
of a few levels just above the threshold energy(9.1 MeV excitation) for the 
separation into $^8$He and $^6$He.

\begin{figure}%----------------------------------------
\noindent
%\begin{minipage}[t]{7cm}
\epsfxsize=0.8\textwidth
\centerline{\epsffile{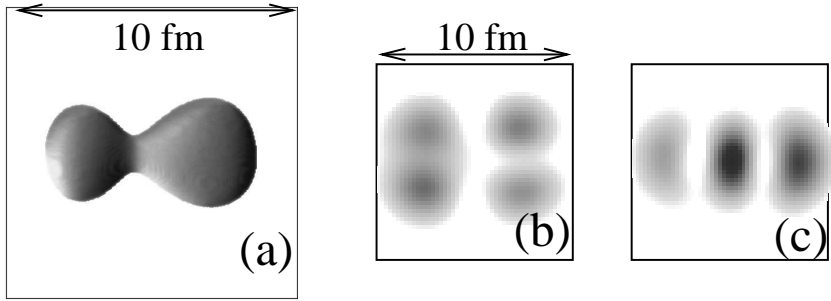}}
%\end{minipage}
%\begin{minipage}[b]{6cm}
\caption{\label{fig:be14}
Intrinsic structure of $0^+_2$ of $^{14}$Be.
Figure (a) shows a surface cut of the matter density $\rho \ge 0.16$ 
nucleons/fm$^3$. The density is calculated for the
intrinsic AMD wave function before spin-parity projection.
Density distributions for the single particle orbits of 
the valence neutrons are presented in the figures (b) and (c).
}
%\end{minipage}
\end{figure}

 The excited states in the region $J\le 6$ 
are classified into 3 rotational bands,
$K=0^+_1$, $K=0^+_2$ and $K=2^+$.
In most of the states in $^{14}$Be, $2\alpha$ cores are formed in the 
calculated results.
As a result, $0^+_1$ and $0^+_2$ states have prolate deformations 
with the deformation parameters
$\beta=0.49$ and $\beta=0.64$ respectively.
By analyzing the single particle wave functions of the intrinsic states,
we found that all the states in the $K=0^+_1$ and $K=2^+$ bands 
are dominated by the normal $0\hbar\omega$ configurations.
What is interesting is the exotic clustering 
structure of the excited states 
in the  second $K^\pi=0^+_2$ band which comes from $2\hbar\omega$ 
configurations.
 In the present calculations, this band
is predicted to start
 from the second $J^\pm=0^+_2$ state at about 
5 MeV and to reach the $8^+_2$ state. 
They have the well-developed $^8$He and $^6$He clusters which are 
very neutron-rich nuclei themselves. 
The surface cut for the matter density
$\rho\ge$ 0.16 nucleons/fm$^3$ of the intrinsic state of 
$0^+_2$ state is shown in Fig.\ref{fig:be14}(a). 
The $^8$He($^6$He) cluster is clearly seen in the right(left)-handed side
in Fig.\ref{fig:be14}(a) where the surrounding low density region is omitted.
As seen in the figure, the spatial clustering develops
remarkably. It is interesting that 
we can find an origin of the clustering in the single particle 
wave functions. According to the analysis of the energies and behaviors of 
the single particle wave functions, the highest 4 neutron orbits correspond
to $sd$-like orbits in the deformed system.
Roughly speaking, a pair of spin up and down neutrons occupies
a higher spatial $sd$-like orbit and the other pair of neutrons occupies
the lower spatial orbit.
Fig.\ref{fig:be14}(b) and Fig.\ref{fig:be14}(c) shows the two types of the 
spatial orbits for the 4 valence neutrons which contain more than 80 \% 
components of positive parity states. 
It is found that the 4 neutrons occupy the $sd$-like orbits modified 
in the prolately deformed system. If we call the longitudinal 
direction of the prolate deformation as $z$-axis,
the highest nucleon orbit seen in Fig.\ref{fig:be14}(b) 
is associated with the $sd$ orbit with a form of $yz \exp [-\nu r^2]$ which 
has nodes for the rotation at the $x$-axis. 
On the other hands, the spatial orbit for the lower one
is similar to the orbit of
$z^2 \exp[-\nu r^2]$ which has nodes along a $z$-axis (Fig.\ref{fig:be14}(c)).
These neutron orbits are stabilized due to the development clustering 
structures because they gain their kinetic energies of the nodes 
along the $z$-axis.
From such a viewpoint of the single neutron orbits we 
can consider that four neutrons in $sd$-like orbits and 2 protons
in $p$-like orbits forms a $^6$He cluster out of a $^8$He core.
In another word, the extremely developed 
$^6$He cluster is originated from a new-type correlation among 4 neutrons 
in $sd$ shell and 2 protons in $p$ shell.

In the theoretical results of $^{15}$B, we find an exotic
clustering $^9$Li+$^6$He structure in the states in the second 
$K^\pi=3/2^-_2$ band which starts from the $J^\pm=3/2^-_2$ 
state about 5 MeV higher than the ground state.
The clusters of $K^\pi=3/2^-_2$ band are also exotic nuclei $^9$Li and
$^6$He, and develop as remarkably as those in the 
$K^\pi=0^+_2$ band of $^{14}$Be.
The mechanism of the $^6$He clustering development in $^{15}$B
can be also described as the correlation among 4 neutrons in the $sd$-shell
 with 2 protons in the $p$-shell as similar way as $^{14}$Be.
That is to say that these excited states with $^9$Li+$^6$He clustering
are based on the $4p-2h$ state of the neutron $p$-shell.
The spatial orbits of 4 valence neutrons are very similar
as those of $^{14}$Be presented in Figs \ref{fig:be14}(b) and 
\ref{fig:be14}(c). 
The correlation among 4 neutrons with 2 protons is very unique
even though the correlation among two neutrons and two protons 
has been sometimes discussed in relation with an $\alpha$ cluster.
In this paper, we suggest one of the novel feature
that 4 neutrons in the $sd$-shell
 with 2 protons in the $p$-shell correlate to form a developed
$^6$He cluster in the neutron rich nuclei. 
  It is an exciting problem whether or not the $^6$He cluster formation
may appear in  other neutron-rich nuclei.

Since the present results are obtained 
within the bound state approximation,
particle decay width has not been discussed though they are important
for the stability of the excited states.
Newly predicted states in $K=0^+_2$ of $^{14}$Be are expected to decay into 
the $^{8}$He+$^{6}$He channel because of the developed clustering.
We have calculated the partial decay width of $^{14}$Be only for
the $^{8}$He+$^{6}$He channel with a method of reduced width amplitude, 
and found that the widths of $6^+_2$ and $8^+_2$ states are less than 50 keV.
For more detailed analysis of the total width including neutron decays and 
excited He decays, we require other microscopic 
frameworks such as a complex scaling method.

\section{Shape coexistence of $^{36}$Ar, $^{40}$Ca}\label{sec:sd}

In the previous section, structures of light neutron-rich have
been discussed. Also in neutron-rich $sd$-shell nuclei we know many 
interesting phenomena such as vanishing of neutron magic number $N=20$ 
in $^{32}$Mg. For study of the structures of neutron-rich nuclei, 
it is very useful to make systematic study associating 
unstable nuclei with stable nuclei.
For example, let us consider a problem of $p$-shell nuclei,
the magic number $N=8$ disappearance in $^{12}$Be.
The intruder ground state is considered to have 2 particles and 2 
holes in neutron $p$-shell. We have already known the similar neutron
configuration in the $0^+_2$ state
of $^{16}$O which is described in terms of $4p-4h$ state with 
$^{12}$C+$\alpha$ clustering. It is very interesting to
imagine the structure change when the system varies from 
$^{16}$O($0^+_2$) to $^{12}$Be with decrease of proton number of the system. 
In case of $sd$-shell nuclei, we think that it is helpful to compare 
the structures of stable nuclei near $^{40}$Ca with those of $^{32}$Mg 
to solve the mechanism of neutron magic number vanishing in neutron-rich
$sd$-shell nuclei. Fortunately many rotational bands of $^{36}$Ar and $^{40}$Ca
have been discovered with the recent $\gamma$-ray measurements.
In this section, the study of the excited states of 
$sd$-shell nuclei $^{36}$Ar and $^{40}$Ca is presented. 
We discuss the shape coexistence problem of these nuclei which is one of
the recent hot subjects.

For the structure study of $sd$-shell nuclei such as $^{36}$Ar and $^{40}$Ca,
we make variation after parity projection but no spin projection (VBP) 
with AMD under a constraint on the total oscillator 
quanta as mentioned in Sec.\ref{sec:formulation}.
The interactions with a finite-range 3-body force are adopted because
the ordinary effective forces such as the MV1 force and the Gogny force 
are not appropriate
to describe the binding energies and radii of nuclei covering wide mass 
number region from $\alpha$ to $^{40}$Ca simultaneously.
The central part of present interactions are as follows,

\begin{eqnarray}
&V_{central}=\sum_{i<j}V^{(2)}+\sum_{i<j<k}V^{(3)},\\
&V^{(2)}=(1-m+bP_\sigma-hP_\tau-m P_\sigma P_\tau)
\left\lbrace 
V_a \exp[-({r_{12}\over r_a})^2]
V_b \exp[-({r_{12}\over r_b})^2]\right\rbrace \\
&+V_c \exp[-({r_{12}\over r_c})^2],\\
&V^{(3)}=V_d \exp[-d(r_{12}^2+r_{23}^2+r_{31}^2)^2],\\
&V_a=-198.34\ {\rm MeV}, V_b=300.86\ {\rm MeV}, V_c=22.5\ {\rm MeV}, \\ 
& r_a=1.2\ {\rm fm}, r_b=0.7\ {\rm fm} , r_c=0.9\ {\rm fm},V_d=600\ {\rm MeV}, d=0.8\ {\rm fm}^{-2}\\
& m=0.193, b=-0.185, h=0.37,
\end{eqnarray}
where the ranges and strength parameters are chosen so as to reproduce 
reasonably the sizes of $\alpha$ and $^{40}$Ca, and the binding energies of 
$\alpha$, $^{16}$O and $^{40}$Ca. To choose interaction parameters 
the other important features like $\alpha$+$\alpha$ phase shift and 
saturation property of the symmetric nuclear matter have been taken into
consideration. In the total interaction, the spin orbit force of G3RS
with the strength $u_{ls}=2500$ MeV and coulomb force are added 
to the central force.

First the excited states of $^{36}$Ar are studied
by VBP calculations with a constraint on 
the total oscillator quanta ${\cal N}$. Here $\Delta {\cal N}$ 
is defined by the deviation from the minimum oscillator quanta of the system; 
$\Delta {\cal N}={\cal N}-{\cal N}_{\rm min}$, where ${\cal N}_{\rm min}$ is
52 in case of $^{36}$Ar.
After spin projection, we obtain an energy surface as a function of 
$\Delta {\cal N}$. It is found that there is a minimum point in the 
energies of $0^+$ states at $\Delta {\cal N}=1$, which corresponds 
to the ground state with 
a normal configuration. By diagonalyzing with the ground state, we obtain
the excited state with $4\hbar\omega$ configurations at $\Delta {\cal N}=4$
as a local minimum state.
As a result, some rotational bands are constructed. In the ground $K=0^+$ band 
the intrinsic state has a normal oblate deformation. On the other hand
the largely deformed state with $4\hbar\omega$ configurations
makes excited rotational bands $K=0^+$ and $K=2^+$. 
In spite of the highly excited configurations the excited $K=0^+$ band 
starts from low energy region about 5 MeV and reaches high spin states 
$J\ge 16$. This excited $K=0^+$ band corresponds well to the experimental
data observed recently with $\gamma$-ray transitions measurements
\cite{SVENSSON}. The intrinsic state of the excited band deforms prolately
as $\beta=0.3$. It is interesting that another shape may coexist 
at $\Delta {\cal N}=5$. Namely, the triaxial shape with 
pentagon component appears to make another
$J^\pm=10^+$ state at about 10 MeV
in present calculations.

\begin{figure}%----------------------------------------
\noindent
\begin{minipage}{9cm}
\epsfxsize=1\textwidth
\centerline{\epsffile{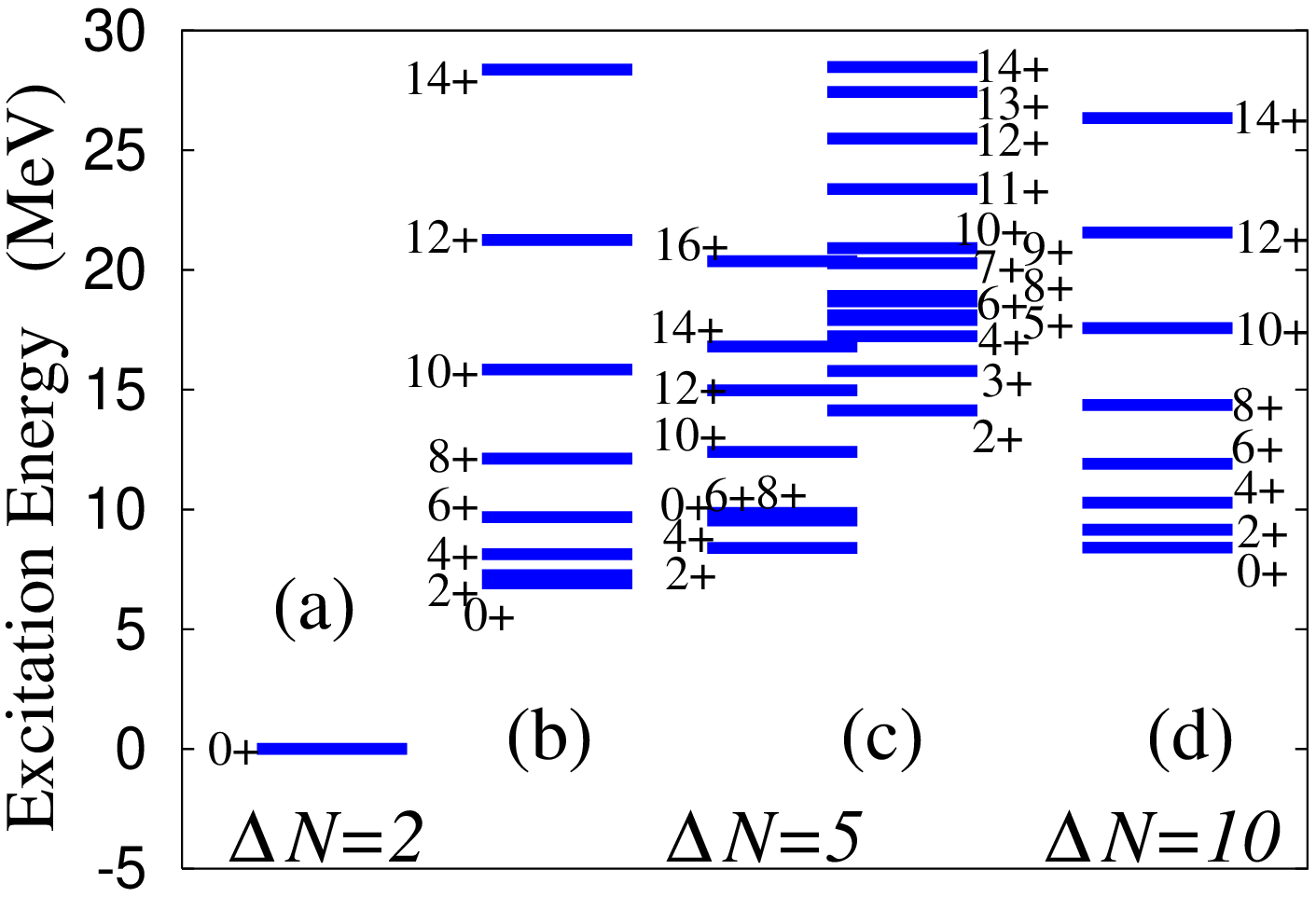}}
\caption{\label{fig:ca40spe}
Theoretical results of energy levels of positive parity states of $^{40}$Ca.
}
\end{minipage}
%\end{figure}
\begin{minipage}{5cm}
%\begin{figure}%----------------------------------------
\noindent
\epsfxsize=1.0\textwidth
\centerline{\epsffile{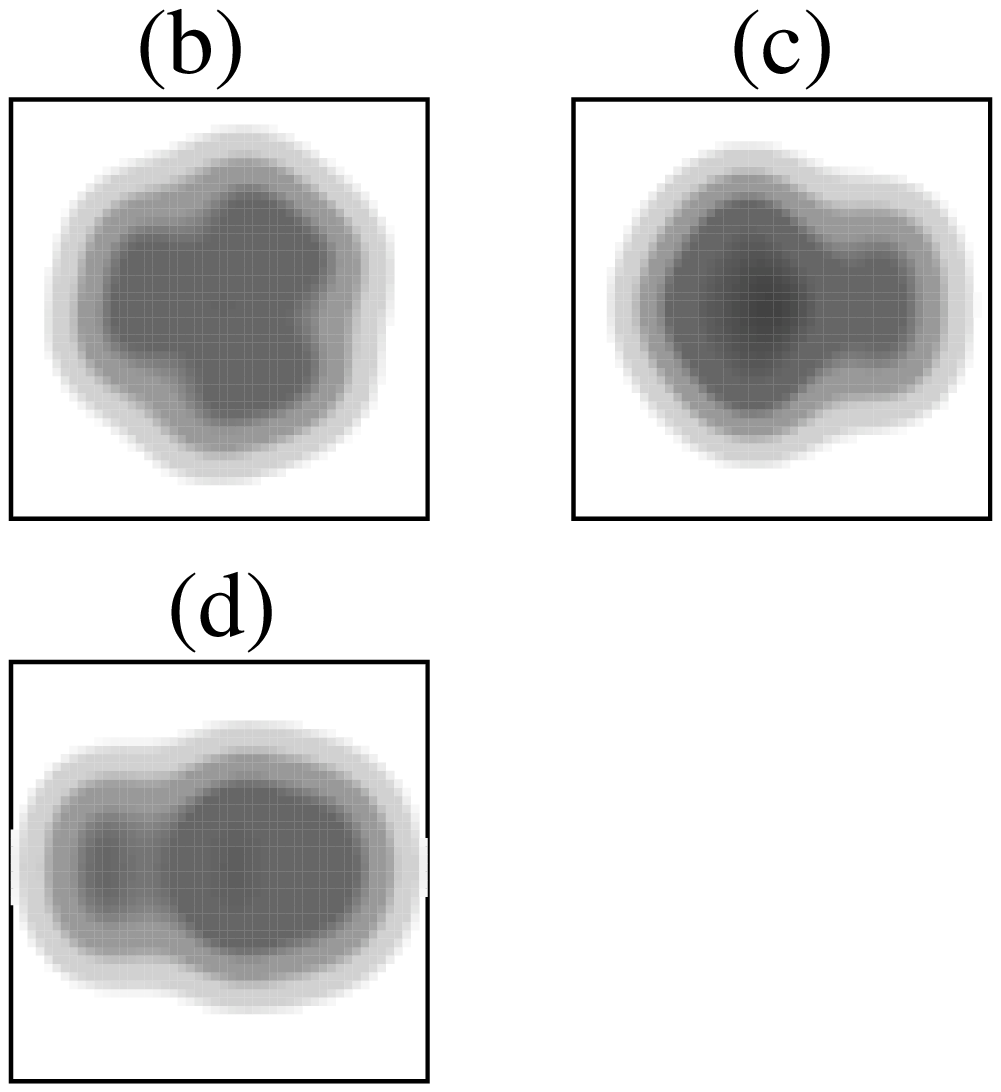}}
\caption{\label{fig:ca40dens}
Density distributions of the intrinsic states of the rotational bands.
Figures (b), (c) and (d) correspond to the oblate shape, the prolate shape
and the super deformation, respectively.
}
\end{minipage}
\end{figure}

In case of $^{40}$Ca, the calculations suggest that 
various kinds of shapes coexist in low energy region although 
$^{40}$Ca is a doubly magic nucleus.
First we find
the ground state with $0\hbar\omega$ configuration at $\Delta{\cal N}=2$.
It has spherical shape (a) as is expected. 
With the increase of oscillator quanta $\Delta{\cal N}$, 
we obtain energetically stable states with $4\hbar\omega$ configurations
at $\Delta{\cal N}=5$. One has an oblate shape (b) and the other has 
a prolate shape (c). Both intrinsic states construct rotational bands
as shown in Fig.\ref{fig:ca40spe}. In the figure only the excitation 
energies of the states with $J\le 16$ are presented because of the limitation
of computational calculations of spin projection. 
As the value $\Delta{\cal N}$ increases, finally a super deformation (d)
appears at $\Delta{\cal N}=10$. By analyzing the single particle wave 
functions
the super deformed state is found ot be dominated 
by $8\hbar\omega$ configurations.
The deformation parameter is $\beta\sim 0.4$.
It is surprising that the rotational band $K=0^+$ 
given from the super deformation starts from the low energy region about 8 MeV
excited in spite of the highly excited configurations of the doubly 
magic nucleus. 
As a result, many shapes coexist in $^{40}$Ca.
It is suggested that the spherical ground state, the oblate excited state, 
the normal prolate excited state, and the prolate super deformed state 
appear in low energy region.
Actually there exist many rotational bands
in the experimental data measured recently \cite{IDEGUCHI}. 
The calculated super deformed band is considered to
be the recently discovered rotational band from 5.273 MeV in the experimental
data. Although the level spacing of the super deformed band 
is overestimated by present calculations, the results should be improved by 
extended calculations such as cranking methods and GCM calculations 
along the constraint $\Delta {\cal N}$.

In the intrinsic states of the excited states,
very strange shapes are found. Figure \ref{fig:ca40dens} shows 
the density distributions of the intrinsic states of the excited bands.
A hexagon like three leaves clover shape appears
In the oblate state (b), while
the super deformed state (d) has a parity asymmetric shape like a pear.
One of the origins of such exotic shapes is clustering effects due to 
$^{12}$C clusters.

As mentioned above, many exotic shapes in the positive parity
states are suggested in the present calculations. 
The appearance of a pear-like 
shape is an exciting prediction because the existence of parity 
asymmetric shapes in such heavy nuclei as $^{40}$Ca is an open problem. 
In order to confirm the parity asymmetric shapes
it is very important to find parity doublet states in negative parity states.
As is expected we find the 
parity doublet states of the super deformation with
a pear shape.
Also in the negative parity states, many shapes are found in the 
theoretical results. We consider the obtained $K=0^-$ at 
$\Delta{\cal N}\approx 8$ is the parity doublet of 
the super deformed band (d) because the intrinsic state is similar to each
other. The excitation energy of the band head $1^-$ state is 12 MeV
which is about 3 MeV higher than the energy of $0^+$ state in the super 
deformation band.
Present results indicate the possibility of the parity asymmetry shape 
due to clustering effects in heavier system.

\section{Summary}

In summary, we have studied the structure of the excited states of $^{12}$Be,
$^{14}$Be and $^{15}$B with variation after spin-parity projection 
based on the framework of AMD method. We have also 
reported our recent study of shape coexistence problems in $^{36}$Ar and
$^{40}$Ca with a constraint AMD. We have obtained a variety of 
exotic structures concerning with clustering.

Various clustering structures in the excited states have been found
in the theoretical results in which we have succeeded to reproduce 
well the excitation energies of many levels of $^{12}$Be.
The breaking of the neuron magic number $N=8$ in 
$^{12}$Be has been found. The recently measured excited states
are described by the third $K^\pi=0^+_3$ band which has a 
foreign clustering structure with developed $^6$He and $^6$He clusters. 
Exotic clustering structures such as $^8$He+$^6$He and $^9$Li+$^6$He have
been predicted in $^{14}$Be and $^{15}$B, respectively.
The formation of the remarkable $^6$He clusters can be understood 
by an unique idea 
of the new type correlation among 4 neutrons in $sd$-shell and 
2 protons in $p$-shell.
This is the first full microscopic calculation
which shows the exotic clustering in the excited states
of $^{12}$Be, $^{14}$Be and $^{15}$B. The results suggest an important
feature that the exotic clustering structures may exist very often
in the excited states of neutron-rich nuclei.

Structures of $^{36}$Ar and $^{40}$Ca have been studied.
In the results it is suggest that many kinds of shapes coexist in
these nuclei. In $^{40}$Ca, the spherical ground state, the normal prolate 
state, the oblate shape, and the prolately super deformation have 
been found.
The prolate state and oblate state are dominated by $4\hbar\omega$ 
configurations, while the super deformation originates from 
$8\hbar\omega$ configurations.
It is surprising that the rotational bands made of such highly excited 
configurations start at low excitation energies.
The rotational band of the super deformation corresponds to the experimentally
measured band which were observed recently in $\gamma$ ray transitions.
In the theoretically obtained intrinsic states, very strange shapes are 
predicted in the excited states. It is very interesting that 
parity asymmetric shape 
like a pear is proposed in the super deformation. 
Although the negative parity bands have not been confirmed experimentally yet,
present results predict that the parity doublet band with negative parity may 
exist at about 3 MeV higher than the positive parity band.

If we vary the system from $^{40}$Ca to $^{32}$Mg by 
decreasing the proton number,
how do the structures of the excited states of stable nuclei change ?
Does the excited state with $4p-4h$ in the neutron shell appear in $^{32}$Mg ? 
It is a very interesting subject to study structures of intruder states in
neutron-rich nuclei associating with the excited of stable nuclei.

\acknowledgements{
I would like to thank Prof. H. Horiuchi for many discussions.
I am also thankful to Dr. N. Itagaki, Prof.
W. Von Oertzen and Prof. Y. Akaishi for helpful discussions 
and comments. Valuable comments of Prof. S. Shimoura and A. Saito
are also acknowledged.
The computational calculations of this work are supported by 
RCNP in Osaka University, YITP in Kyoto University and
IPNS/KEK.}

%\section*{References}


\begin{thebibliography}{9}
  
\bibitem{ENYOf}
 Y. Kanada-En'yo, H. Horiuchi and A. Dot\'{e},
J. Phys. G, Nucl. Part. Phys. {\bf 24} 1499 (1998).
\bibitem{OGAWA}
Y. Ogawa, K. Arai, Y. Suzuki, and K. Varga, Nucl. Phys. {\bf A673}
122 (2000).
\bibitem{DOTE}
A. Dot\'{e}, H. Horiuchi, and Y. Kanada-En'yo, 
Phys. Rev. C {\bf 56}, 1844 (1997).
\bibitem{ITAGAKI}
N. Itagaki and S. Okabe, Phys. Rev. C {\bf 61}, 044306 (2000).
\bibitem{ENYOg}
Y. Kanada-En'yo, H. Horiuchi and A. Dot\'{e} Phys. Rev. {\bf C 60}, 
064304(1999)
\bibitem{FREER}
M. Freer, et al., Phys. Rev. Lett. {\bf 82}, 1383 (1999).
\bibitem{TSUZUKIa}
T. Suzuki and T. Otsuka, Phys. Rev. {bf C} 56, 847(1997).
\bibitem{IWASAKI}
H. Iwasaki, et al., Phys.Lett.{\bf B 491}, 8(2000).
\bibitem{ENYObc}
 Y. Kanada-En'yo, H. Horiuchi and  A. Ono,
Phys. Rev. C {\bf 52}, 628 (1995);
 Y. Kanada-En'yo and H. Horiuchi,
Phys. Rev. C {\bf 52}, 647 (1995).
\bibitem{ENYOe}
 Y. Kanada-En'yo,
Phys. Rev. Lett. {\bf 81}, 5291 (1998).
\bibitem{ENYOsup}
 Y. Kanada-En'yo and H. Horiuchi, to be published in Prog. Theor. Phys. Suppl..
\bibitem{TOHSAKI}
 T. Ando, K.Ikeda, and A. Tohsaki, Prog. Theor. Phys.
 {\bf 64}, 1608 (1980).
\bibitem{LS}
 N. Yamaguchi, T. Kasahara, S. Nagata, and Y. Akaishi,
 Prog. Theor. Phys. {\bf 62}, 1018 (1979);
 R. Tamagaki, Prog. Theor. Phys. {\bf 39}, 91 (1968).
\bibitem{ITAGAKIa}
N. Itagaki, S. Okabe and K. Ikeda, Phys. Rev. C {\bf 62}, 034301 (2000).
\bibitem{TANIHATA}
 A.A. Korsheninnikov, et al., Phys. Lett. B343, 53 (1995).
\bibitem{OERTZEN}
W. von Oertzen, Z. Phys. A {\bf 354}, 37 (1996).
\bibitem{SAITO}
 A. Saito and S. Shimoura, private cominucations.
\bibitem{SVENSSON}
C.E.Svensson et al., Phys.Rev.{\bf C63}, 061301 (2001)
\bibitem{IDEGUCHI}
E.Ideguchi et al.,PHys.Rev.Lett.{\bf 87}, 222501(2001)
\end{thebibliography}
\end{document}